\documentclass[%
 reprint,
 superscriptaddress,
 amsmath,amssymb,
 aps,
]{revtex4-2}

\usepackage{xr}
\usepackage{graphicx}
\usepackage{epstopdf}
\usepackage{dcolumn}
\usepackage{bm}
\usepackage{bbm}

\newcommand{\xv}{\mathbf{x}}

\newcommand{\pv}{\mathbf{p}}

\newcommand{\Sv}{\mathbf{S}}

\begin{document}


\title{Fisher information for Markovian Trajectories of Gene Expression}
\title{Extracting Information from Stochastic Trajectories of Gene Expression}

\author{Zachary R Fox}
\email{zachfox@lanl.gov}
\affiliation{Center for Nonlinear Studies (T-CNLS), Theoretical Division, Los Alamos National Laboratory, Los Alamos NM 87545
}%
\affiliation{Information Sciences Group (CCS-3), Computer, Computational and Statistical Sciences Division, Los Alamos National Laboratory, Los Alamos NM 87545}
 
\date{\today}

\begin{abstract}
Gene expression is a stochastic process in which cells produce biomolecules essential to the function of life. Modern experimental methods allow for the measurement of biomolecules at single-cell and single-molecule resolution over time. Mathematical models are used to make sense of these experiments. The codesign of experiments and models allows one to use models to design optimal experiments, and to find experiments which provide as much information as possible about relevant model parameters. Here, we provide a formulation of Fisher information for trajectories sampled from the continuous time Markov processes often used to model biological systems, and apply the result to potentially correlated measurements of stochastic gene expression. We validate the result on two commonly used models of gene expression and show it can be used to optimize measurement periods for simulated single-cell fluorescence microscopy experiments. Finally, we use a connection between Fisher information and mutual information to derive channel capacities of nonlinearly regulated gene expression.
\end{abstract}

\maketitle

As the revolution of single-cell biology has revealed the importance of fluctuations in understanding how gene are regulated \cite{Elowitz:2000,Munsky:2012Science,Raj:2006}, biomeasurement technologies have been developed to precisely measure individual molecules relevant to all components of transcription, translation, and regulation \cite{Bertrand:1998, Larson:2011,Morisaki:2016}.
Even under stationary conditions, biomolecule abundances fluctuate over time, and the covariances and correlations contain useful information about the underlying biophysical parameters of the system \cite{Munsky:2012}. 
Much experimental advancement in recent years is microscopy methods to measure \textit{trajectories} of biomolecule abundances in individual cells over time, yet historically such systems are analyzed using moment-based and/or continuous approximations \cite{Chait:2017, Ruess:2015}, or assume temporal independence between measurements \cite{Neuert:2013,Golding:2017,GomezSchiavon:2017}.
However, many such biomolecular systems operate in the low-copy number limit, where there are effects on the discrete number of molecules and non-negativity. 
Because much relevant gene regulation occurs within this regime, continuous approximations of the dynamics are not appropriate and can lead to inaccurate inference and predictions \cite{Munsky:2018}. 
Here, we develop a computational method based on stochastic path integrals and information theory to design single-cell experiments using computational models without resorting to moment-based approximations or temporal-independence assumptions.

Fisher information has become a promising mathematical construct for designing single-cell experiments \cite{Komorowski:2011,Ruess:2013,Fox:2019, Fox:2020} and for an information theoretic understanding of how cells process their environments \cite{Mora:2015, Jetka:2018, Mora:2019, Vennettilli:2021}.
Current approaches for Fisher information of stochastic gene regulation either assume temporal independence \cite{Fox:2019, Fox:2020} or approximate master equation dynamics with Langevin equations that assume Gaussian fluctuations in time and state \cite{Komorowski:2011,Ruess:2013,Haas:2013, Zimmer:2016}. 
Here we introduce the Fisher information for stochastic sample paths of continuous time Markov chains (CTMCs) and demonstrate how it can be applied to stochastic gene expression models. 
The path-based information calculation provides correct analyses when temporal correlations are present in the data, and in some sense is the natural formalism to analyze gene expression trajectories.
\begin{figure}[h!]
    \centering
    \includegraphics[width=\columnwidth]{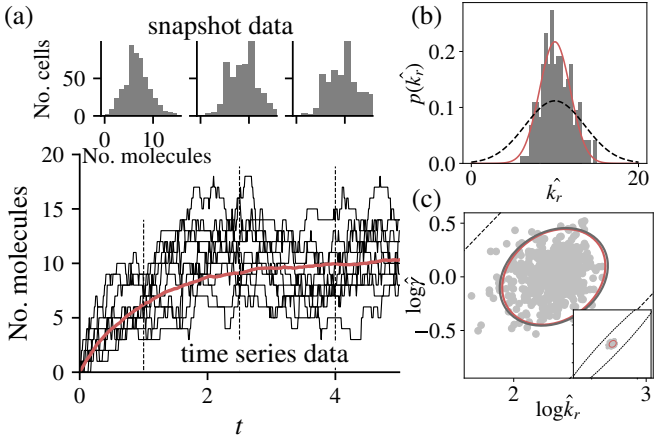}
    \vspace{-.25cm}
    \caption{ Application of the Fisher information for stochastic trajectories of a birth-death process. (a) Trajectory data follow individual cells over time, shown in black. This contrasts snapshot data, which are temporally independent as the individual cells are killed to take measurements. (b) Distribution of maximum likelihood estimates for the single parameter $k_r$ for 200 individual trajectories are shown in gray, and as predicted by the trajectory based Fisher information, $\mathcal{I}^{ts}$ (red). The birth rate $k_r=10$, decay rate $\gamma=1$, and trajectories were sampled every $\Delta t=0.26$ time units, for a total of 20 time points. The black dashed line shows the predicted information calculated by assuming snapshot measurements $\mathcal{I}^{ss}$. (c) Same as in (b), except for two parameters $k_r$ and $\gamma$, and using 200 time points instead of 20. The red ellipse indicates the 95\% CI for the FSP-FIT, and the grey ellipse is estimated from the MLE estimates. The inset with the black dashed line shows the expected uncertainty from snapshot data.} \label{figs:fig1}
\end{figure}
CTMC's are used throughout the physical sciences \cite{Verley:2016}, epidemiology \cite{Allen:2017} and mathematical finance \cite{Turra:2016}. Each state in the CTMC can be written as vector $\mathbf{x}_i = \left[ \zeta_1, \zeta_2, \dots, \zeta_N\right]_i \in \mathbf{X} \subset \mathbb{Z}^N_{\geq 0}$. Transitions from state $\xv_i - \psi_{\nu}$ to $\xv_i$ occur in the infinitesimal time $dt$ with probability $w_{\nu}(\zeta,\theta,t) dt$, where $\theta$ refers to the kinetic parameters of the process, and $\psi_{\nu}$ is a vector which dictates the $\nu^{\rm th}$ state transition. These rates can be assembled into a (potentially infinite) generator matrix $\mathbf{A}$, and the transition probabilities $\mathbf{P}_{ji}(\Delta t)$ of hopping from $\mathbf{x}_i$ to $\mathbf{x}_j$ in the time $\Delta t$ can be found by integrating the autonomous set of ODEs $\dot{\mathbf{P}} = \mathbf{A} \mathbf{P}$ from $t=0$ to $t=\Delta t$. 
For simplicity in notation, let us consider a single-species model, such that $\xv_i = [\zeta_1]_i = x_i$.

Let us define a trajectory sampled from the CTMC at regular intervals of $\Delta t$, $\vec{X}_t = [x(0), x(\Delta t), ..., x( N_t\Delta t)]$, as this type of sampling is common in time-lapse microscopy. For example, $\vec{X}(t)$ may correspond to measurements of the number of protein molecules in a given cell every $\Delta t$ time units. 
For simplicity, we will use $x(k \Delta t) \equiv x_k$ throughout the manuscript. 
Using the Markov property, the log-probability of a given trajectory is
\begin{align} \label{traj_prob}
    \log P(\vec{X}_t)  &= \log p(x_0) + \sum_{k=1}^{N_t}\log p(x_k|x_{k-1}). 
\end{align}
In this work, we use  Eq.\ \ref{traj_prob} to find the Fisher information for the kinetic model parameters, denoted $\theta$. 
The Fisher information is defined
\begin{align}\label{FIM_path}
    \mathcal{I}_{i,j} = \mathbb{E} \left[ \frac{\partial \log P(\vec{X}_t)}{\partial \theta_i}\frac{\partial \log P(\vec{X}_t)}{\partial \theta_j} \right],
\end{align}
where the expectation is taken over the distribution of possible paths \cite{Kay:1993, Haas:2013}. 
We find the CTMC based Fisher information for trajectories to be 
\begin{align} \label{FIT_FIM}
        \mathcal{I}^{ts}_{i,j} &= \underbrace{\mathbbm{1}^{\rm T} \left( \frac{1}{\pv_0} \odot \frac{\partial}{\partial \theta_i} \pv_0 \odot \frac{\partial}{\partial \theta_j} \pv_0\right)}_{\text{information in initial distribution}} \nonumber \\
        & +  \underbrace{\mathbbm{1}^{\rm T}\left( \sum_{k=0}^{N_t-1} \mathbf{Z}^{(\alpha \beta)} \mathbf{P}^k \right) \pv_0}_{\text{information in state transitions}}.
\end{align}
where $\mathbf{Z}^{(\alpha \beta)}_{ij}=\Sv^{\theta_i}_{lm}\Sv^{\theta_j}_{lm} / \mathbf{P}_{lm}$ and $\odot$ indicates elementwise multiplication, and the vector $\pv_0$ is the initial distribution over the states.  These results are derived in Supplement \ref{sec:FIM-deriv} using a straightforward path-integral approach.
Note that this formulation does not require the system to be stationary. 
To compute $\mathcal{I}^{ts}$ Eq.\ \ref{FIM_path}, the sensitivities of $\mathbf{P}$ to the parameter $\theta_i$, $\mathbf{S}^{\theta_i}$ can be found using forward sensitivity analysis, similar to that in \cite{Fox:2020,Fox:2019}, by solving the following set of coupled ordinary differential equations,
\begin{align} \label{trans_pdf}
    \frac{d}{dt}\begin{bmatrix} \mathbf{P} \\ \mathbf{S}^{\theta_i} \end{bmatrix} = 
    \begin{bmatrix}
    \mathbf{A} & \mathbf{0} \\
    \mathbf{A}^{\theta_i} & \mathbf{A} 
    \end{bmatrix}
    \begin{bmatrix} \mathbf{P} \\ \mathbf{S}^{\theta_i} \end{bmatrix}.
\end{align}
A derivation for Eq.\ \ref{trans_pdf} is provided in Supplement \ref{sec:MJP-sens}.
Equation \ref{FIT_FIM} gives the expected information from a single measured trajectory. If one measures $N_c$ independent trajectories under the same experimental conditions, the information simply scales linearly, i.e. $\mathcal{I}_{\rm total} = N_c \mathcal{I}^{ts}$ \cite{Kay:1993}. 

In general, Fisher information matrices can be numerically verified using the asymptotic normality of the maximimum likelihood estimator, 
\begin{align} \label{asymp_norm}
    \sqrt{N_c} \left( \theta^{*}-\hat{\theta} \right) \xrightarrow{\rm dist} \mathcal{N}(\mathbf{0}, \mathcal{I}^{-1}).
\end{align}
In other words, under mild regularity conditions \cite{Kay:1993} the distribution of maximum likelihood estimates of model parameters $p(\hat{\theta})$, given sufficient data, must converge to a normal distribution with a covariance matrix given by the inverse Fisher information (i.e. the Cram\'er-Rao bound) \cite{Casella:1990}.

Now, we consider the process of stochastic gene expression and regulation, in which cells create mRNA through a process called transcription, and these mRNA are translated into protein in a process called translation. The production and degradation/dilution of mRNA may be regulated in many ways leading to complex, nonlinear and non-Gaussian dynamics.
When these processes are described within the CTMC framework above, the governing set of ODEs for the probability of each state is referred to as the chemical master equation (CME) \cite{VanKampen:1992,McQuarrie:1967}. Each state in the CTMC is a vector of the integer counts of each molecular species. Within the parlance of the CME framework, $\psi_{\nu}$ is referred to as the stoichiometry vector, as it determines the integer change of $\mathbf{x}$ for the $\nu^{\rm th}$ biochemical reaction, and $w_\nu (\zeta, \theta, t)$ refers to the propensity of a given biochemical reaction. 
With knowledge of a particular gene regulatory circuit, these transitions can be written into a (potentially) infinite generator matrix $\mathbf{A}$.

We first study the trajectory based FIM for the classic birth-death system, which can be used to model the production and degradation of a single mRNA or protein \cite{Zenklusen:2008, Munsky:2012Science}. The system consists of two reactions, production and degradation of some molecule $X$, which occur with rates  $\theta=[k_r,\gamma]$, 
\begin{align}
    \varnothing \xrightarrow{k_r} X; \hspace{.5cm} X \xrightarrow{\gamma} \varnothing 
\end{align}
The infinitesimal generator is given in Supplement \ref{sec:CT-BD}.
Figure \ref{figs:fig1}(a) shows sample trajectories of the birth-death process, as well as histograms of snapshot measurements at three time points. 

To verify the trajectory based information $\mathcal{I}^{ts}$ in Eq. \ref{FIT_FIM}, we use the asymptotic normality of the maximum likelihood estimator in Eq.\ \ref{asymp_norm}. We simulated 200 trajectories of the process using the stochastic simulation algorithm (SSA) \cite{Gillespie:1977}. The production rate $k_r=10$ and decay rate $\gamma =1$, and the measurements were taken every $\Delta t = 0.26$ time units, for a total of 20 measurements.
To generate maximum likelihood estimates for each trajectory, we apply a simple Nelder-Mead optimization algorithm to find the combination $\hat{\theta}$ which maximizes Eq.\ \ref{traj_prob}. 
The most simple case is with a single free parameter, i.e. $\theta=[k_r]$. We show the distribution of MLEs of $k_r$ as the gray histogram in Fig.\ \ref{figs:fig1}(b). The red line shows a normal distribution with variance given by $1/\mathcal{I}(k_r)$. We then applied the same numerical experiments when $\theta=[k_r,\gamma]$ using 200 time points, and again show excellent agreement between theory (red) and simulation (grey) in Fig.\ \ref{figs:fig1}(c). Each ellipse corresponds to a 95\% confidence interval for the normal distribution. 
\begin{figure}[h!]
\centering
\includegraphics[width=\columnwidth]{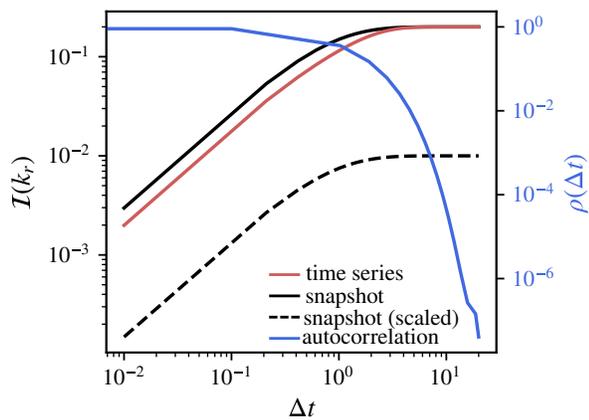}
\vspace{-1.2cm}
\caption{Information about transcription rate $k_r$ as a function of $\Delta t$, using the trajectory based calculation (red), and the snapshot based calculation (black solid line). The blue line and right axis show the expected correlation of the number of molecules as $\Delta t$ increases. The dashed black line shows the expected snapshot information when it is corrected for the number of measurements that are required to ensure independent sampling.}\label{figs:fig2}
\end{figure}

Next, we investigate how the information of time series measurements $\mathcal{I}^{ts}$ compares to the Fisher information matrix for snapshots developed in \cite{Fox:2019,Fox:2020},
\begin{align} \label{FIM_ss_nc}
        \mathcal{I}^{ss}_{i,j} = \sum_{k=1}^{N_t} N_c(t_k) \sum_{l=1}^N \frac{1}{p(x^{(l)}_k)}\partial_{\theta_i}p(x^{(l)}_k)\partial_{\theta_j}p(x^{(l)}_k).
\end{align}
 In snapshot measurements, independent subpopulations are population are sampled independently in time and therefore there are no temporal correlations between the measurements. 
If measurements are close together in time, they will be correlated, and therefore statistically less valuable than a set of independent measurements, i.e. it has a lower effective sample size.  
However, it is possible that the correlations themselves are sensitive to model parameters, and correlation information may improve parameter estimation. Therefore the trajectory based Fisher information is not a lower bound on the snapshot information \cite{Komorowski:2011}. 
We study the difference between snapshot and time-series measurements for the birth-death model described above by varying the time between measurements, $\Delta t$.
Figure \ref{figs:fig2} shows snapshot information exceeds time-series information for the single-measurement limit (black and red solid lines, respectively), indicating that there is not enough information in correlations to overcome the statistical independence of the snapshot measurements. As expected, when $\Delta t$ becomes large, time-series measurements are no longer correlated, and the snapshot information and correlated information converge to the same value. The autocorrelation of molecule number $\rho(t)$ decays as the $\mathcal{I}^{ss}$ and $\mathcal{I}^{ts}$ converge (blue curve in Fig.\ \ref{figs:fig2}, where $\frac{d}{dt}\rho(t)=\phi \rho(t)$, $\rho(0)=1.0$ and $\phi$ is the autonomous ODEs for the system \cite{Gardiner:2004}. However, in the correlated regime, the number of cells being measured, $N_c$, is necessarily more for the snapshot data, as one must measure independent trajectories at each time point to obtain statistically independent data. In other words, the number of measured trajectories must be proportional to the number of time points $N_c \propto N_t$. Therefore, the accurate comparison of information provided per trajectory can be found by taking $\mathcal{I}^{ss}/N_t$, and is shown by the black dashed line in Fig.\ \ref{figs:fig1}(b-c) and Fig.\ \ref{figs:fig2}. 

\begin{figure}[ht!]
\centering
\includegraphics[width=.8\columnwidth]{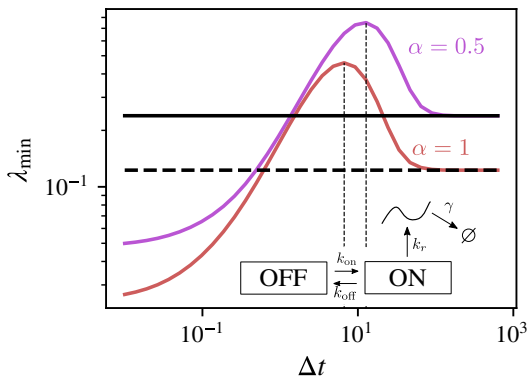}
\caption{ Experimental optimization of the measurement period $\Delta t$ for a simple two-state gene expression model. The system starts at a stationary distribution, and thus the snapshot information is independent of $\Delta t$ (black horizontal lines). However, information in the correlation with a given trajectory changes as a function of measurement period (red and purple lines). The units of time in this example are minutes. $\lambda_{\rm min}$ corresponds to the smallest eigenvalue of the information matrix. }\label{figs:fig3}
\end{figure}

Next we consider a model of an unregulated gene that stochastically switches between an active state, in which gene products mRNA are made, and an inactive state in which they are not made, as shown in Fig.\ \ref{figs:fig3}. This model of gene switching has been used extensively in the literature to describe stochastic gene expression \cite{Peccoud:1995,Shahrezaei:2008}, and under different parameterizations is capable of reproducing rich biological phenomena. 
This system consists of four reactions:
\begin{align}
    \mathcal{G}_{\rm off} &\xrightarrow{k_{\rm on}} \mathcal{G}_{\rm on}; \hspace{0.5cm} \mathcal{G}_{\rm on} \xrightarrow{k_{\rm off}} \mathcal{G}_{\rm off} \nonumber \\
    \mathcal{G}_{\rm on} &\xrightarrow{k_{\rm r}} \mathcal{G}_{\rm on} + {\rm mRNA}; \hspace{0.5cm}{\rm mRNA} \xrightarrow{\gamma} \varnothing
\end{align}
We set $k_{\rm on}=0.05 \alpha \text{ min}^{-1}$, $k_{\rm off}=0.15 \alpha \text{ min}^{-1}$, $k_{\rm }=5 \text{ min}^{-1}$, and $\gamma=0.05 \text{ min}^{-1}$, where the parameter $\alpha$ dictates the time scale of the gene switching. 
We first verified $\mathcal{I}^{ts}$ for this model when both the gene state $\{ \mathcal{G}_{\rm off},\mathcal{G}_{\rm on} \}$ and $X$ are measured (see Supplement \ref{sec:two_state}) using the asymptotic normality in Eq.\ \ref{asymp_norm}.

Next, we demonstrate how the measurement period $\Delta t$ can be optimized to learn about the transition rates between gene states $k_{\rm on},k_{\rm off}$ in addition to the mRNA production rate, $k_r$. This serves as a simple example of how one might use the FIM to optimize the camera acquisition rate for a fluorescent time-lapse microscopy experiment, such as those in \cite{Elowitz:2000,Stewart:2012}. 
We compute the Fisher information for $\theta=[k_{\rm on},k_{\rm off},k_r]$ at different values of $\Delta t$ between 0.1 minutes and 315 minutes, and compute the minimum eigenvalue $\lambda_{\rm min}$ of the matrix $\mathcal{I}^{ts}(\theta)$ at each $\Delta t$. This eigenvalue is commonly maximized in optimal experiment design, as it corresponds to the least-informative direction of the parameter space. By maximimizing $\lambda_{\rm min}$, one effectively shrinks parameter uncertainty in the least informative direction.
We assume the system is already at a stationary distribution i.e. $\mathbf{p}_0 = \mathbf{p}_{\rm ss}$, and therefore the snapshot information is constant over $\Delta t$, shown by the black horizontal lines in Fig.\ \ref{figs:fig3}. 
However, the trajectory based information can leverage the sensitivity of the correlation of the process to extract information about the model parameters. At short $\Delta t$, the system does not change much over the measurement horizon and samples are highly correlated, and therefore little information is gained. At long $\Delta t$, measurements are effectively independent and $\mathcal{I}^{ss}$ and $\mathcal{I}^{ts}$ converge, as shown by the black and red/purple lines in Fig.\ \ref{figs:fig3}.
Between these two limits, there is an optimal $\Delta t$, which changes for different switching time scales, $\alpha = [0.5,1]$, as shown by the red and purple lines in Fig.\ \ref{figs:fig3}. The dashed lines indicate the values of $\Delta t$ that optimize this metric of $\mathcal{I}^{ts}$. The optimal value of $\Delta t$ is $12.8$ min for $\alpha=0.5$ and $6.7$ min for $\alpha=1$. One could use this approach to choose the camera acquisition rates for future experiments.
So far, we have focused on the information that stochastic trajectories carry about model parameters. However, biological systems must process information about their environments, and information theoretic approaches have been established to understand the fundamental limits of biochemical sensing and regulatory circuits \cite{Berg:1977, Mora:2019, Razo:2020, Tkacik:2009}. Much work has been performed to calculate mutual information of gene regulatory motifs using myriad approximations \cite{Tkacik:2009, Walczak:2010, Tkacik:2011, Jetka:2018}, though these methods rely on Gaussian approximations, small noise limits, and steady-state approximations. Here, we use a connection between Fisher information and mutual information ($I(X;Y)=H(X)-H(X|Y)$, $H(\cdot)=\sum p(\cdot) \log p(\cdot)$) as developed in \cite{Clarke:1990,Brunel:1998}, to analyze the channel capacity of a regulated gene, taking into account non-Gaussian fluctuations and discrete copy number effects. The channel capacity is of interest as it characterizes the maximum information that the gene circuit can process about its environment. It is typically derived by optimizing the input distribution $p(X)$ to maximize I(X;Y) \cite{Shannon:1948}. In \cite{Jetka:2018,Jetka:2019} the authors used this connection along with the Linear Noise Approximation (instead of the more general Chemical Master Equation approach developed here) to analyze dynamic outputs as well as implicitly analyze dynamic inputs for the IFN signaling network. These previous works show that the channel capacity of each individual cell can be determined using the Fisher information as
\begin{align} \label{chan_cap} 
    \mathcal{C}_k = \log_2 \left( 2\pi e^{\frac{1}{2}} \int_{\mathcal{X}} dx \sqrt{\mathcal{I}(x)} \right),
\end{align}
\begin{figure}[ht!]
\centering
\includegraphics[width=\columnwidth]{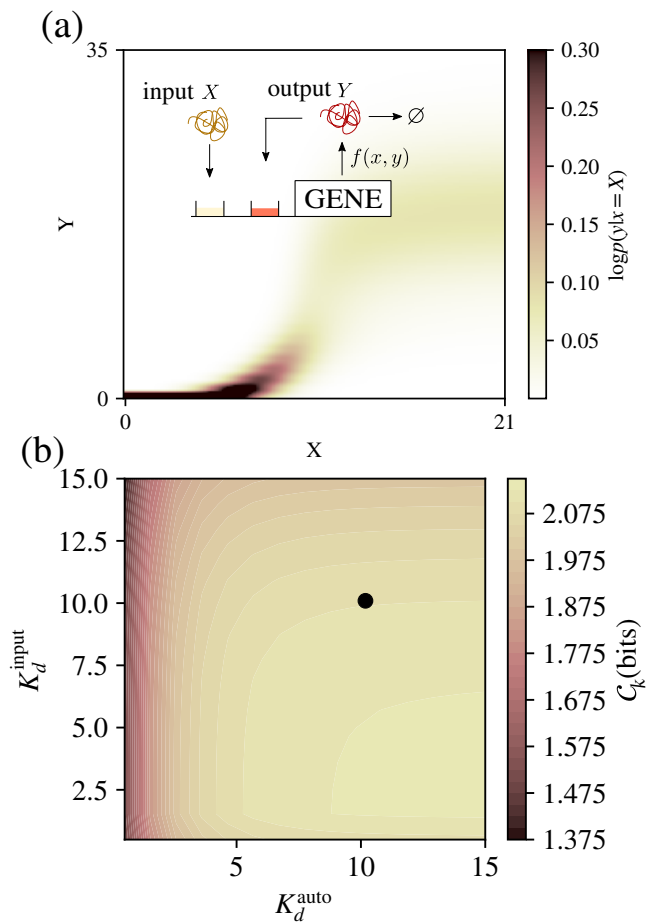}
\caption{Channel capacity of a coregulated gene. (a) Diagram of the input TF (x) and output TF (y) and stationary probability distributions of $Y$ for different values of $X$ shown. (c) Channel capacity of the circuit in (a) as a function of two model parameters $K_d^{\rm auto}$ and $K_d^{\rm input}$. }\label{figs:fig4}
\end{figure}
and the total channel capacity for a collection of $N$ cells is given by $\mathcal{C}_N = \mathcal{C}_k + 1/2 \log_2{N}$, i.e. the number of distinguishable inputs $2^{\mathcal{C}_N}$ increases linearly with the number of cells $N$ \cite{Brunel:1998, Jetka:2018}. By applying the CME-based, trajectory based FIM to Eq.\ \ref{chan_cap}, we can compute the channel capacity of a stochastic, nonlinearly regulated circuit without assumptions on the state transition densities. 

We demonstrate this approach with an input-output gene regulatory circuit, in which the input is the concentration of a transcription factor $x$ and the output is a stochastic trajectory $\vec{Y}(t)$, similar to \cite{Jetka:2018, Tkacik:2009,Tkacik:2011}, Fig.\ \ref{figs:fig4}(a). In addition to the regulation of the output by the input, the protein $Y$ is also able to upregulate itself through a positive feedback loop, with OR-type logic, leading to the total nonlinear production term
\begin{align}
    f(x,y) = \frac{k^{\rm input}_r x^n}{(K^{\rm input}_d)^n + x^n}+\frac{k^{\rm auto}_ry^n}{(K^{\rm auto}_d)^n + y^n}.
\end{align}
The output distribution of $p(Y)$ at stationarity is shown in Fig.\ \ref{figs:fig4}(b) for different input values of $X$. Given the values of parameters shown in in Fig.\ \ref{figs:fig4}(a),  at intermediate $X$, the nonlinear activation leads to a bistable system, and individual trajectories may switch stochastically between high and low outputs over time (see Supplement \ref{sec:info_theory_background} for more details). One would imagine that this could give rise to at least one bit of information about $X$ by simply reading out $Y$ as being low or high. In practice, these parameter setting provide $\mathcal{C}_k \approx 2$ bits of information about the input $X$ when we consider the dynamic response (black dot in Fig.\ \ref{figs:fig4}(b)). 
We then evaluated the channel capacity of this circuit by performing a grid search over relevant values of two parameters $K^{\rm auto}_d$ and $K^{\rm input}_d$, and found that the circuit's channel capacities range from $1.40$ to and $2.13$, for trajectories of $10$ measurements with $\Delta t = 0.1 $ time units. These channel capacities are comparable to values seen in the literature \cite{Tkacik:2011}.

We presented a computational analysis of Fisher information matrices for stochastic trajectories measured at discrete times. Compared with previous approaches to time-series data in \cite{Komorowski:2011,Zimmer:2016}, this approach does not rely on Langevin approaches or Gaussian distribution transition densities. Furthermore, this is agnostic to nonlinearities in propensity functions, whereas previous approaches would need to use moment closure schemes to accurately find the higher order statistics of the Langevin approximation. While we have so far considered a simple experiment design variable $\Delta t$, another relevant design variable involves the addition of measurement noise to the system, such as the Poisson-like statistics coming from measurements of single fluorophores, recently coined probabilistic distortion operators \cite{Vo:2021} in the context of Fisher information analyses. Furthermore, when only a subset of species are observed, the likelihood in Eq.\ \ref{traj_prob} can no longer simply make use of the Markov property, as one must marginalize over the unobserved species. Future work will investigate such situations.  

We envision the results in this work find use in the synthetic biology community, in which modern experimental platforms \cite{Rullan:2018, CastilloHair:2019, Chait:2017} allow for feedback between computers and single-cells using optogenetic transcription factors enable online optimal design of parallelized experiments among individuals within a single population of growing and dividing cells. Outside of biology, this approach could in principle be applied to any continuous time Markov chain observed at regular intervals, subject to computational constraints of the method. 

We thank Yen Ting Lin for his helpful discussions and comments. We also thank Brian Munsky, Huy Vo, and Nishant Panda for their feedback on this Letter. ZRF gratefully acknowledges the support of the U.S. Department of Energy through the LANL/LDRD Program and the Center for Nonlinear Studies for this work. 

\bibliography{fsp_fim,gen}

\end{document}